\documentstyle[12pt]{article}
\def\beq{\begin{equation}}
\def\eeq{\end{equation}}
\def\bea{\begin{eqnarray}}
\def\eea{\end{eqnarray}}

\def\ba{\begin{array}}
\def\ea{\end{array}}
\def\bce{\begin{center}}
\def\ece{\end{center}}

\def\nonu{\nonumber}

\begin{document}
\begin{titlepage}
\rightline{SNUTP-98-021}
\rightline{UM-TG-206}
\rightline{hep-th/9803197}
\def\today{\ifcase\month\or
January\or February\or March\or April\or May\or June\or
July\or August\or September\or October\or November\or December\fi,
\number\year}
\vskip 1cm
\centerline{\Large \bf  Comments on $SO/Sp$ Gauge Theories from}
\centerline{\Large  \bf Brane Configurations
with an O6 Plane}
\vskip 1cm
\centerline{\sc Changhyun Ahn$^{a}$,
Kyungho Oh$^{b}$
and Radu Tatar$^{c}$}
\vskip 1cm
\centerline{{\it $^a$ Dept. of Physics, Seoul National University,
Seoul 151-742, Korea}}
\centerline{{ \it $^b$ Dept. of Mathematics, University of Missouri-St.
Louis,
St. Louis, MO 63121, USA}}
\centerline{{\it $^c$ Dept. of Physics, University of Miami,
Coral Gables, FL 33146, USA}}
\vskip 2cm
\centerline{\sc Abstract}
\vskip 0.2in
We use the M theory approach in the presence of an orientifold O6 plane
to understand some aspects of the moduli space of vacua for $N=1$
supersymmetric $SO(N_c)/Sp(N_c)$ gauge theories in four dimensions.
By exploiting some general properties of the O6 orientifold,
we reproduce some results obtained previously with an orientifold O4 plane
when the flavor group arises from the worldvolume dynamics of D6 branes.
By using semi-infinite D4 branes instead of D6 branes, we derive the most
general form of the
rotated curve describing the moduli space of vacua
for $N=1$ supersymmetric gauge theory.
\vskip 1in
\leftline{revised September 1998}
\end{titlepage}
\newpage
\setcounter{equation}{0}

\section{Introduction}
\setcounter{equation}{0}

Recently, important progress has been made in studying
the strongly coupled dynamics of  low energy  supersymmetric gauge theories
in
various dimensions.
The D-brane dynamics provides a powerful geometrical description for
the gauge theories which are obtained on their world-volume.
A detailed account of
many aspects of the interrelation between D-brane dynamics and
supersymmetric gauge theory in different dimensions can be found in the
very interesting review of Giveon and Kutasov \cite{gk}.

Many aspects of the strongly coupled dynamics of supersymmetric gauge
theories
have been explained by using string theory results.
The mirror symmetry of the $N=4$ gauge theory in 3
dimensions was described in \cite{hw} as being due to the nonperturbative
S-duality of type IIB string theory (See also \cite{bo1}).
A stringy derivation of Seiberg's duality for $N=1$ supersymmetric
$SU(N_c)$
gauge theory with flavors appeared in \cite{egk,egkrs}.
This description was generalized to the brane configurations with
orientifolds
where the gauge group is either $SO(N_c)$ or
$Sp(N_c)$ \cite{eva} (See also \cite{ov} for an  equivalent
geometrical  approach).

In string theory (10 dimensions) there are singularities where the branes
are
touching each other. The singularities
are removed in 11 dimensions where the brane configuration becomes
smooth.
Both the D4 branes and NS brane used in type IIA string theory become a
unique
M5 brane in 11 dimensions. The D6 branes are  the Kaluza-Klein monopoles
given by
Taub-NUT space \cite{town}.
The world volume of the M5 brane was observed to be the product of the
four dimensional spacetime and the Seiberg - Witten curve uniquely
identified with the solutions for the Coulomb branch of the four
dimensional gauge theories \cite{w1}.
Further generalizations of this configuration were obtained
by inserting an O4 orientifold \cite{lll}.
The low energy description of $N=1$ supersymmetric $SU(N_c)$ gauge
theories with flavors in 4 dimensions have been found
in \cite{w2,hoo,biksy} (See
also \cite{aotaug,aotsept} for theories with an orientifold 4 plane).

Many results have been very recently obtained from brane configurations
in the presence of an O6 plane \cite{egkrs,ll,lll1,bhkl,egkt,csst,uranga}.
The brane configuration is very similar with the one constructed with an
orientifold O4 plane but it allows us to obtain new gauge theories on the
world-volume of the D4 branes lying between two NS branes like
theories with $SO\times  SU \times SU...$ gauge groups \cite{ll} or
theories with matter in symmetric and antisymmetric representations
\cite{lll1,bhkl,egkt,csst,uranga}.

In this paper, we study the M5 brane with an orientifold O6 plane,
along the lines of \cite{w2,hoo,biksy} in order to
understand the moduli space of vacua of $N=1$ supersymmetric
$SO(N_c)/Sp(N_c)$ gauge theories in 4 dimensions.
We will see how our previous results of
\cite{aotaug,aotsept} are rederived in the presence of an O6 orientifold
instead of an O4 orientifold. We also consider the case of semi-infinite
D4 branes instead of D6 branes and explicitly derive the form of the
M-theory curve for the rotated NS branes configuration by using the method
developed in \cite{biksy}.  We obtain as solution for
the result constructed in \cite{csst}.

In section 2 we briefly review the results obtained in \cite{ll} for the
$N=2$ configuration. In section 3 we consider the case when the flavor
group
is given by D6 branes and we rederive the results of \cite{aotaug,aotsept}
in the presence of O6 orientifold. In section 4 we explicitly derive the
M-theory curve for rotated configuration in the presence of semi-infinite
D4 branes.

\section{Brane Configuration for $N=2$}
\setcounter{equation}{0}

We would like to review the work of
Landsteiner and Lopez \cite{ll} where they considered
brane configurations of type IIA string theory
giving $N=2$ $SO(N_c)$ or $Sp(N_c)$ gauge theory.
In order to study the orthogonal and symplectic groups,
an O6 plane parallel to the D6 branes is added into the
$SU(N_c)$ gauge theory brane configuration. The O6 plane
does not break further supersymmetry and there exist two possible signs for
the O6 plane charges.
The brane configuration  consists of
NS5 branes, D4 branes and D6 branes together with an orientifold
O6 plane. Working on the double covering of the orientifold,
we denote their worldvolumes:
\bea
NS5 & : & \;\;\; (x^0, x^1, x^2, x^3, x^4, x^5) \nonu \\
D4  & : & \;\;\; (x^0, x^1, x^2, x^3, x^6) \nonu \\
D6/O6 & : & \;\;\; (x^0, x^1, x^2, x^3, x^7, x^8, x^9).
\eea
Here the D4 branes suspended between two NS5 branes
are  finite in $x^6$ direction.
O6 plane acts as a
mirror in $(x^4, x^5, x^6)$ directions due to the spacetime reflection
and
the two NS5 branes are mirror images
of each other under this orientifold projection.
Every D4 branes which does pass through $x^4=x^5=0$ should have its mirror
image.

As usual
we write two complex coordinates as
$v=x^4+i x^5, s=(x^6+i x^{10})/R, t=e^{-s}$
where $x^{10}$ is the 11th coordinate of M theory which is compactified
on a circle of radius $R$.

$\bullet$ $SO(2N_c)$ Case

Let us first construct the M theory curve for the $SO(2N_c)$ gauge theory
in the presence of O6 plane.
The O6 plane in $SO$ gauge theory carries a
$+4$ D6 brane charge. Thus the singularity associated with the O6-plane
can be expressed as a quotient of a surface
\bea
xy =  v^4
\label{v4}
\eea
in ${\bf C}^3$ by   a ${\bf Z_2}$
symmetry $ x \leftrightarrow y$ and $v \leftrightarrow -v$ which
corresponds to a $D_4$ singularity as observed in \cite{ll,witten3}.
Moreover, in the presence of D6 branes, the surface (\ref{v4})
 in ${\bf C}^3$ will be
generalized to
\bea
xy = v^4 \prod_{i=1}^{N_f}(v^2 -m_i^2)
\label{complex}
\eea
where the orientifold projection
allows only the configuration
invariant under the $ x \leftrightarrow y$ and $v \leftrightarrow -v$.
In summary, M-theory for $SO$ gauge group in the presence of
O6 plane and D6 branes should be  described on
the quotient space of $xy = v^4 \prod_{i=1}^{N_f}(v^2 -m_i^2)$ by
a ${\bf Z_2}$
symmetry $ x \leftrightarrow y$ and $v \leftrightarrow -v$ replacing
the ordinary flat spacetime  ${\bf R^3} \times \bf S^1$ (the coordinates
being $x^4, x^5,x^6, x^{10}$).

The directions corresponding to (\ref{complex}) are transversal to the
D6 branes.
The D6 branes are located at $x=y=0, v=\pm m_i$ and $N_f$ is the
number of ``physical'' D6 branes.

Let us describe how the above brane configuration
appears in M theory as a
a generically smooth single M5 brane whose worldvolume has,
in addition to four spacetime dimensions, another 2 directions given by
$\Sigma$.
Furthermore, we can identify $\Sigma$  with
the Seiberg-Witten curve \cite{aps} which determines  the solution
of the  Coulomb branch of the gauge theory.
Following \cite{w1}, the Seiberg-Witten curve $\Sigma$ can be described by
\bea
& & xy = \Lambda_{N=2}^{4N_c-4-2N_f} v^4 \prod_{i=1}^{N_f}(v^2 -m_i^2)\\
& & x +y = B(v^2).
\eea
where $B$ is an even polynomial of degree $2N_c$ in $v$.
We introduced the power of $\Lambda_{N=2}$ in order to match the
dimensions.
The projected image of this curve  in $(y,v)$-space will be given by
\bea
y^2 - B(v^2) y +
\Lambda_{N=2}^{4N_c-4-2N_f} v^4 \prod_{i=1}^{N_f}(v^2 -m_i^2) = 0.
\eea
It is easy to see that by redefining $\tilde{y}=y/v^2$ this
reproduces the form of the curve of \cite{aps}.

$\bullet$ $SO(2N_c+1)$ Case

In this case we have a supplementary D4 brane which has no mirror
and is stuck at $v=0$. Then the Seiberg-Witten curves are:
\bea
& & xy = \Lambda_{N=2}^{4N_c-2-2N_f} v^4 \prod_{i=1}^{N_f}(v^2 -m_i^2)\\
& & -x +y = v B(v^2)
\eea
with the same orientifold projection as before.
The Seiberg-Witten curve  will be projected to
\bea
y^2 - v B(v^2) y -
\Lambda_{N=2}^{4N_c-2-2N_f} v^4 \prod_{i=1}^{N_f}(v^2 -m_i^2) = 0.
\eea
in $(y, v)$-space
and one can check that this leads to the result of \cite{aps}
by redefining $\tilde{y}=-y/v^3$.

$\bullet$ $Sp(N_c)$ Case

In case of the symplectic gauge group,
O6 plane carries a $-4$ D6 brane  charge.
The orientifold is described by
\bea
xy = v^{-4}.
\eea
with the same orientifold projection as before.
As before, the presence of D6 branes modify the previous equation to
\bea
xy = v^{-4} \prod_{i=1}^{N_f}(v^2 -m_i^2).
\eea
with the same orientifold projection as before.
The curve is then given by
\bea
& & xy = \Lambda_{N=2}^{4N_c+4-2N_f} v^{-4} \prod_{i=1}^{N_f}(v^2 -m_i^2)\\
& & x +y = C(v^2)+\Lambda_{N=2}^{2N_c+2-N_f} v^{-2}.
\eea
with the same orientifold projection as before.
Away from the orientifold, the curve  is described by
a single equation in $(y,v)$-space
\bea
y^2 - ( C(v^2)  + \Lambda_{N=2}^{2N_c+2-N_f} v^{-2}
\prod_{i=1}^{N_f} m_i ) y+
\Lambda_{N=2}^{4N_c+4-2N_f} v^{-4} \prod_{i=1}^{N_f}(v^2 -m_i^2) = 0
\label{swequ}
\eea
Note that $C(v^2)$ is a polynomial in $v$ of degree $2N_c$ and the
existence
of $v^{-2}$ due to the orientifold projection dominates for small $v$ while
it can be ignored for large $v$.
If we write  $\tilde{y}$ as $v^2 y$ the previous relation becomes
the one in \cite{aps}.

\section{Rotated Configuration with D6 Branes }
\setcounter{equation}{0}

In this section we are going to work only within the brane configuration
in order to
derive the moduli space of vacua for the corresponding $N=1$ theory.
We are not going to derive the results in field theory which were obtained
in
detail in \cite{aotaug,aotsept} (see also \cite{bhoo})
and we will limit ourselves to some comments at
specific points. We refer the interested reader to
\cite{aotaug,aotsept} for the derivation of the field theory results.

$\bullet$ $SO(2N_c)$ Case

It is convenient to introduce a complex coordinate $w=x^8+i x^9$.
Before breaking the $N=2$ supersymmetry, the two NS5 branes are located
at $w=0$.
Now we rotate the right NS5 brane towards $w$ direction and this determines
its mirror
image, the left NS5 brane, to be rotated towards its mirror direction.
This  gives a mass to the adjoint chiral multiplet in the $N=2$ vector multiplet. In order to rotate these NS5 branes, all the D4 branes are placed together
and the motion of the D4 branes along $v$ direction is not possible, 
Here we are assuming that the  $N=2$ curve is irreducible.
The Higgsing is
possible only when the D4 branes breaks on the D6 branes. (This is an interesting
case.) For this reason,
we will assume that the D6 branes passes through $v=0$ i.e. we turn off the bare mass $m_i$. In short, we are rotating the following curve:
\bea
\label{curve}
x +y = v^{2N_c}, \,\,\, xy = \Lambda_{N=2}^{4N_c-4-2N_f} v^{2N_f +4}.
\eea
The asymptotic behavior of the rotated right NS5 brane and left NS5 brane
when $v \rightarrow \infty$ imposes
 the following boundary conditions for $SO(2N_c)$
\bea
\label{bdy-cond}
& & w \rightarrow \mu v \;\;\; \mbox{as}\;\;\; v \rightarrow
\infty, \;\;\; y \sim v^{2N_c}  \nonumber \\
& & w \rightarrow -\mu v \;\;\; \mbox{as}\; \;\; v \rightarrow
\infty, \;\;\; y \sim
\Lambda_{N=2}^{2(2N_c-2-N_f)}v^{2N_f-2N_c+4}
\eea
in the threefold $V$ defined by 
\bea
\label{3fold}
xy =\Lambda_{N=2}^{4N_c-4-2N_f} v^{2N_f +4}.
\eea
Thus the worldvolume of the M5 brane describing the $N=1$ $SO(2N_c)$ gauge theory will be given by ${\bf R}^{3,1} \times \Sigma$, where the
algebraic curve $\Sigma$ fits the boundary
conditions $(\ref{bdy-cond})$ and is embedded in the threefold $V$.
We will describe the projection of $\Sigma$ 
 into $(y,v,w)$-space.  
First we can compactify the curve $\Sigma$  by adding two points
at infinity $v \rightarrow \infty$. 
that the functions $w_+ \equiv w +\mu v$ or $w_- \equiv w-\mu v$ have a
simple
pole at one of the points
at infinity. This implies that $\Sigma$ is a rational curve which can be
globally parameterized by either $w_+$ or $w_-$.
Thus we can express the functions  $y$ and $v$ on $\Sigma$
in terms of $w_+$ by rational function.
\bea
\label{PQ+}
v=P(w_+), \;\;\;\;\; y=Q(w_+).
\eea
Since the  orientifold projection sends $v \to -v$ and $y \to x$,
the following equation must hold:
\bea
\label{PQ-}
v=-P(w_-), \;\;\;\;\; x=Q(w_-).
\eea
Since $v$ and $y$ are finite except at $w_+ =0, \infty$, these
rational functions are polynomials of $w_+$ up to a factor of
some power of $w_+$: $P(w_+) = w_+^{a}p(w_+)$ and $Q(w) =
w_+^{b}q(w_+)$ where
$a$ and $b$ are some
integers and $p(w_+)$ and $q(w_+)$ are polynomials of
$w_+$ which we may assume
nonvanishing at $w_+=0$. Near one of the points at $w_+=\infty$,
$v$ and $y$ behave
as $v\sim \mu^{-1}w_+$ and
$y \sim v^{2N_c }$ by (\ref{bdy-cond}). Thus the rational
functions are of the form
\bea
P(w_+) = w_+^{a}(w_+^{1-a} + \cdots)/2\mu\;\;\; \mbox{and}\;\;\;
Q(w_+) = \mu^{-2N_c}w_+^{b}(w_+^{2N_c -b} + \cdots)
\label{PQ}
\eea
Now around the other infinity $w_+ =0$,
the curve $\Sigma$ can be parameterized
by $1/v$ which vanishes as
$w_+\to 0$ from the boundary condition.
Since $w_+$ and $1/v$ are two coordinates around the neighborhood $w_+=0$
in the  compactification of $\Sigma$
and vanish at the same point, they must be linearly proportional to
each other
$w_+ \sim 1/v$ in the
limit $w_+ \to 0$.
Thus the function $P(w_+)$ takes the form
\bea
\label{P}
P(w_+)=\frac{w_+^2 + \cdots }{2\mu  w_+} =
 \frac{(w_+ - w_1)(w_+ - w_2)}{2\mu w_+}.
\eea
\bea
\label{P+=P-}
\frac{(w_+ - w_1)(w_+ - w_2)}{\mu w_+} =
-\frac{(w_- - w_1)(w_- - w_2)}{\mu w_-}.
\eea
Now by putting $w=0$ in this equation, we obtain
\bea
2(w_1 +w_2)\mu v =0.
\label{wequa}
\eea
Since $w_+ \sim 1/v$ in the
limit $w_+ \to 0$, the functions $v$ and $w$ can not vanish simultaneously
on $\Sigma$.
Hence we have $w_1 = - w_2$. We let $w_0 =w_1$. Now the equation (\ref{P})
becomes
\bea
\label{v}
v= P(w_+) = \frac{(w_+^2 - w_0^2)}{2\mu w_+}.
\eea
Since  $y \sim v^{2N_f-2N_c+4}$ and $w_+ \sim 1/v$ as $w_+\to 0$, we get
$b=2 N_c -4  -2N_f$ and thus,
\bea
\label{y}
y=Q(w_+) = \mu^{-2N_c }w_+^{2N_c-4 -2N_f}(w_+^{2N_f +4 } + \cdots ).
\eea
By substituting $(\ref{v})$ and $(\ref{y})$ into $(\ref{3fold})$,
we conclude
\bea
\label{Q}
y=Q(w_+)=\mu^{-2N_c} w_+^{2N_c-4-2N_f} (w_+^2 - w_0^2)^{N_f
+2}.
\eea
Note that the solutions $(\ref{v})$ and $(\ref{Q})$ satisfy
the conditions $(\ref{PQ-})$. In fact, we have
\bea
w_- = w_+ - 2\mu v = w_+ -2\mu P(w_+) = w_0^2/w_+
\eea
and 
\bea
P(w_0^2/w_+) = -P(w_+) \;\;\;\;\; 
Q(w_0^2/w_+) = x (\mbox{up to a constant factor})
\eea
as required by  $(\ref{PQ-})$.

To find the values of  $w_0$, we observe that
the solutions $v =P(w_+)$
and $y = Q(w_+)$ should satisfy $(\ref{curve})$ namely
\bea
y^2 - v^{2N_c }y +\Lambda_{N=2}^{4N_c-4-2N_f} v^{2N_f+4} = 0.
\eea
in order to keep the $U(1)$ symmetry in the $w$ direction as pointed out
in \cite{hoo}.
By plugging the equations (\ref{v}) and (\ref{Q}) in the previous equation
this equation and equating the lowest order terms in $(w_+ \pm w_0)$,
we obtain :
\bea
w_0=(-1)^{\frac{N_f}{4N_c-4-2N_f}} \mu \Lambda_{N=2} \;\;\;\; \mbox{or}
\;\;\;\;
w_0=0.
\eea

As described in \cite{hoo}, the values for $w$ are just the expectation
values
for the meson. The above results are identical with the ones obtained in
\cite{aotsept} both in field theory (where the values for $w$ are
the expectation values for the meson field $M$) and in
brane configuration with an O4 plane. We refer to \cite{aotsept} for the
physical interpretation of the result.

Without matter, the curve describing the $N=2$ Coulomb branch is given by
\bea
y^2 - B(v^2) y +
\Lambda_{N=2}^{4N_c-4} v^4  = 0.
\eea
The genus of the curve is $2N_c-1$ provided $B(v^2)$ is a general
polynomial.
By dividing $v^4$ and renaming $y v^{-2} $ as $ \tilde{y} $, this becomes
\bea
\tilde{y}^2 - B(v^2) \tilde{y} v^{-2} +
\Lambda_{N=2}^{4N_c-4}  = 0.
\eea
Then this curve is completely degenerate at $2N_c-2$ points on the Coulomb
branch. At one of these points, the curve has the following form
\bea
v^2=\tilde{y}^{\frac{1}{2N_c-2}}+\Lambda_{N=2}^{2} \tilde{y}^{-\frac{1}{2N_c-2}}.
\eea

$\bullet$ $Sp(N_c)$ Case

For this case we can proceed in a very similar way.
We just describe the facts without much details.
The boundary conditions from the rotations of two NS5 branes
can be read off easily by looking the behavior of $y$ and $v$ in the
(\ref{swequ}) as follows:
\bea
\label{bdy-cond1}
& & w \rightarrow \mu v \;\;\; \mbox{as}\;\;\; v \rightarrow
\infty, \;\;\; y \sim v^{2N_c}  \nonumber \\
& & w \rightarrow -\mu v \;\;\; \mbox{as}\; \;\; v \rightarrow
\infty, \;\;\; y \sim
\Lambda_{N=2}^{2(2N_c+2-N_f)}v^{2N_f-2N_c-4}.
\eea
Since the functions  $y$ and $v$ on $\Sigma$
can be written in terms of $w_+$ as rational function similar to
(\ref{PQ+})
and the extra conditions (\ref{PQ-})
arising from the orientifold
action hold as well, by following the same procedure through
(\ref{PQ}) and (\ref{wequa}) we will get
\bea
P(w_+) = \frac{(w_+^2 - w_0^2)}{2\mu w_+}.
\label{Psp}
\eea
Since  $y \sim v^{2N_f-2N_c-4}$ and $w_+ \sim 1/v$ as $w_+\to 0$, we get
$b=2 N_c +4  -2N_f$. This leads to
\bea
\label{Qsp}
y=Q(w_+)=\mu^{-2N_c} w_+^{2N_c+4-2N_f} (w_+^2 -w_0^2)^{N_f -2}.
\eea
The solutions $v =P(w_+)$
and $y = Q(w_+)$ should satisfy
by
\bea
y^2 - v^{2N_c }y +\Lambda_{N=2}^{4N_c+4-2N_f} v^{2N_f-4} = 0.
\eea
By plugging (\ref{Psp}) and (\ref{Qsp}) in
this equation and taking the limit $w_+ \to  \pm w_0$,
we obtain
\bea
w_0=(-1)^{\frac{N_f}{4N_c+4-2N_f}} \mu \Lambda_{N=2}.
\eea
This result is again in accordance with the results obtained in
\cite{aotaug} in field theory and from brane configuration in the presence
of
an O4 plane. We refer to \cite{aotaug} for the physical interpretation of
the
results.

Without matter, the curve describing the $N=2$ Coulomb branch is given by
\bea
y^2 - C(v^2) y +
\Lambda_{N=2}^{4N_c+4} v^{-4}  = 0.
\eea
The genus of the curve is $2N_c-1$ provided $C(v^2)$ is a general
polynomial.
By multiplying $v^4$ and renaming $y v^{2} $ as $ \tilde{y} $, this becomes
\bea
\tilde{y}^2 - C(v^2) \tilde{y} v^{2} +
\Lambda_{N=2}^{4N_c+4}  = 0.
\eea
Then this curve is completely degenerate at $2N_c+2$ points on the Coulomb
branch. At one of these points the curve has the following form
\bea
v^2=\tilde{y}^{\frac{1}{2N_c+2}}+\Lambda_{N=2}^{2} \tilde{y}^{-\frac{1}{2N_c+2}}.
\eea
\section{Rotated Configuration with Semi-infinite D4 Branes }
\setcounter{equation}{0}

$\bullet$ $SO(2N_c)$ Case.

Now  we insert $N_f$ semi-infinite D4 branes extending to the left
from  the left
NS5 brane, and their mirrors extending to the right from the right NS
brane.
The equation for the Sieberg-Witten curve is given by
\bea
xy& =&\Lambda_{N=2}^{4N_c -4 -2N_f} v^4\\
\label{w}
w& =& 0 \\
\label{x+y}
(-1)^{N_f} \prod_{i=1}^{N_f}(v + m_i) x &+&
\prod_{i=1}^{N_f}(v - m_i) y = B(v^2)
\eea
with orientifold projection $x \leftrightarrow y$ and
$v\leftrightarrow -v$.
 One observation is in
order here: in the case of an O4 plane, the variable $v$ always appears
in equations as $v^{2}$ because for
 each D4 brane at $ v$ there is one at $-v$.
The O6 plane introduces the symmetry $x^{6} \leftrightarrow - x^{6}$
besides
the $v \leftrightarrow - v$ symmetry. So the effect on semi-infinite D4
branes
is the following: each semi-infinite D4 brane ending at $v$ on the left of
the
left NS brane has a mirror ending at $- v$ on the right of the right NS
brane.
For this reason in equation (\ref{x+y})
we have terms with $ v$ and not $v^{2}$.

Since for large $y$ with fixed or small $x$, $y$ corresponds to $t$ and
for large $x$ with fixed or small $y$, $x$ corresponds to $t^{-1}$,
the equation (\ref{x+y}) becomes
\bea
&&\prod_{i=1}^{N_f}(v - m_i) y  \sim  B(v^2),\quad \mbox{for large } y\\
&&\prod_{i=1}^{N_f}(v +  m_i) x \sim  B(v^2)
,\quad \mbox{for large } x.
\eea
This shows that the above equations describe the brane configuration
under consideration.
The Seiberg-Witten curve will be projected to
a curve  in $(y, v)$-plane given by
\bea
\label{even}
\prod_{i=1}^{N_f}(v -  m_i) y^2 - B(v^2) y +
(-1)^{N_f} \Lambda_{N=2}^{4N_c -4 -2N_f}v^4\prod_{i=1}^{N_f}(v + m_i) = 0.
\eea
Since in a brane configuration of these  models
the Coulomb branch is not modified by a change of the
singular locus,
we may simplify the spacetime   by resolving  the
singularities. We desingularize  the surface $S$,
$xy = \Lambda_{N=2}^{4N_c -4 -2N_f} v^4$, by blowing up
the ideal $(x, y, v^2)$.
Then the resolved surface $S'$ can  be described by
\bea
x'y' = \Lambda_{N=2}^{4N_c -4 -2N_f}
\eea
and the new surface $S'$  will map onto the singular surface $S$ by
$x = x' v^2$ and $y = y' v^2$.  Now the Seiberg-Witten curve on this new
space will be given by
\bea
x'y'& =&\Lambda_{N=2}^{4N_c -4 -2N_f} \\
\label{prime-w}
w &=& 0 \\
\label{prime-x+y}
(-1)^{N_f} \prod_{i=1}^{N_f}(v + m_i)v^2 x'& +&
\prod_{i=1}^{N_f}(v - m_i)v^2 y' = B(v^2)
\eea
with an orientifold projection $x' \leftrightarrow y'$ and
$v\leftrightarrow -v$.
In this new space, the Sieberg-Witten curve will be projected to
a curve in $(y', v)$-plane given by
\bea
\label{prime-even}
\prod_{i=1}^{N_f}(v -  m_i)v^{2} y'^2 - B(v^2) y' +
(-1)^{N_f} \Lambda_{N=2}^{4N_c -4 -2N_f}v^2\prod_{i=1}^{N_f}(v + m_i) = 0.
\eea
In terms of brane geometry, the effect of resolving the singularity
corresponds to pushing  $+4$ charge of O6 through the NS5 branes to
$\pm\infty$, creating new non-dynamical D4 branes  \'a la Hanany-Witten
as described by Uranga~\cite{uranga}.

Now let us  rotate the left NS5 brane towards $w$ which determines
 the rotation of its mirror right brane towards $- w$ direction.
Thus the left  NS5 brane is located at $w_{+} = 0 $ and the right NS5
brane is located at $w_{-} = 0$ where we have again:
\bea
w_\pm = w \pm \mu v.
\eea
We can make another choice of coordinates to describe the brane
configuration
but $w_\pm$ are the most efficient in our approach  as we will see below.
Also we are only dealing in this section
with  massive quarks having  generic masses
 $m_1, \ldots , m_{N_f}$ and we realize this in the brane configuration
by putting the semi-infinite D4 branes at $(v,w) =( -m_i/2\mu, m_i/2)$
on the left NS5 brane and their mirrors at $(v,w) =(m_i/2\mu, m_i/2)$
 on  the right NS5 brane.
Let $\Sigma$ be the corresponding Seiberg-Witten curve.
On $\Sigma$, the function $w_+$ goes to infinity only at one point and
$w_+$ has only a single pole there, since there is only one NS5 brane i.e.
the right NS5 brane. Thus we can identify $\Sigma$ with the
punctured complex $w_+$-plane possibly after resolving the singularity
at $x=y=v=0$.
Similarly, we can argue that $w_-$ has a single pole on $\Sigma$.
 Since these are two rational functions on a rational curve,
they are related by
a linear fractional transformation which after suitable constant shifts
can be written as
\bea
\label{w_+w_-}
w_+w_- = \zeta
\eea
where
$\zeta$ is a constant.

Now we project this curve to $(y, w_+)$-space to obtain:
\bea
\label{w_+}
\prod_{i=1}^{N_f}(w_+ - m_i) y  - P_{1}(w_+) = 0,
\eea
where
\bea
 P_{1}(w_+) = w_+^{2N_c} + p_1  w_+^{2N_c -1 } + \cdots + p_{2N_c}
\eea
is some polynomial of degree $2N_c$ because there are $2N_{c}$ finite D4
branes
between the two NS branes and the the number of finite D4 branes gives the
degree of $P_{1}$.
Similarly if we project the curve to $(y, w_-)$-space, we get
\bea
\label{w_-}
Q_{1}(w_-) y  - A \prod_{i=1}^{N_f}(w_- -  m_i) = 0
\eea
where
\bea
 Q_{1}(w_-) = w_-^{2N_c} + q_1  w_-^{2N_c -1 } + \cdots + q_{2N_c}
\eea
and $A$ is a normalization constant. In writing the equations (\ref{w_+})
and (\ref{w_-}) we have used the fact that one of the functions
$w_{+}, w_{-}$  vanishes at  one of the
NS branes so when we project the curve to  one of the $(y,w_{\pm})$ spaces,
its
equation becomes linear in $y$. We have used the idea of \cite{biksy} where
the configuration had two NS branes with the left NS brane was
extended in the $v$ direction
at $w = 0$ and the right NS brane was extended in the $w$ direction at
$v = 0$ so the brane equations were linear when projected to $(y, v)$ or
$(y, w)$ spaces.

In order to have (\ref{w_+w_-}), (\ref{w_+}) and (\ref{w_-})
simultaneously,
it is required that
\bea
\label{bisky}
P_{1}(w_+)Q_{1}(\zeta/w_+ )
 \equiv A\prod_{i=1}^{N_f}(w_+ - m_i)(\zeta/ w_+ -  m_i)
\eea
for all $w_+ \in {\bf C}$.
The most general solution for the massive flavor case is:
\bea
\label{csstP}
P_{1}(w_+)& =& w_+^{2N_c -N_f}\prod_{i=1}^{N_f}(w_+ -\zeta / m_{i})\\
\label{csstQ}
Q_{1}(w_-)& =& w_-^{2N_c -N_f}\prod_{i=1}^{N_f} (w_- - \zeta/ m_i).
\eea
Now we plug (\ref{csstP}) into (\ref{w_+}) to obtain a Seiberg-Witten
curve:
\bea
16 \mu^4 xy& =&\Lambda_{N=2}^{4N_c -4 -2N_f} (w_+ - w_-)^4\\
w_+w_- &=& \zeta\\
\label{CSSTlike}
\left(\prod_{i=1}^{N_f} m_i \right)
y \prod_{i=1}^{N_f}\left(\frac{w_+ - m_i}{w_- - m_i}\right) &=&(-1)^{N_f}
w_+^{2N_c}.
\eea
As before, we may push the charge of O6 plane through the NS5 branes
to $\pm\infty$ by going  to the resolved surface $S'$
given by
\bea
\label{resolved}
x'y' = \Lambda_{N=2}^{4N_c -4 -2N_f}.
\eea
Similar computations
can be done on the resolved surface $S'$ after replacing
(\ref{w_+}) and (\ref{w_-}) by
\bea
\label{prime-w_+}
\prod_{i=1}^{N_f}(w_+ - m_i)w_+^{2} y'  - P_{1}(w_+) = 0.\\
\label{prime-w_-}
Q_{1}(w_-) y'  -A w_-^{2}\prod_{i=1}^{N_f}(w_- -  m_i) = 0.
\eea
Then as the most general solution for
massive flavor, we obtain:
\bea
\label{prime-CSSTlike}
\left(\prod_{i=1}^{N_f} m_i \right)
y' \prod_{i=1}^{N_f}\left(\frac{w_+ - m_i}{w_- - m_i}\right) = (-1)^{N_f}
w_+^{2N_c -2}.
\eea
Moreover, from (\ref{prime-w_-}) we can identify A with
$\Lambda_{N=1}^{3(2 N_c -2) - 2 N_f}$ by taking into account the geometric
symmetry $U(1)_{w_{+}} \otimes U(1)_{w_{-}}$ of the M theory associated
with
the complex rotations of the left and right NS branes respectively.
Thus the value for $\zeta$ can be obtained as
\begin{equation}
\zeta = (\Lambda_{N=1}^{3(N_c -1) - N_f}
 \prod_{i=1}^{N_f} m_{i})^\frac{1}{N_{c} - 1}
\end{equation}
by comparing (\ref{prime-w_+}) and (\ref{prime-w_-}) as in (\ref{bisky}).
If we map this curve to $(x, y, w_+, w_-)$-space  via
$x= v^2x',  y=v^2 y'$,
we will obtain  a set of  equations similar to (5.2), (5.3) and  (5.4)
 of \cite{csst}:
\bea
\label{us}
xy& =&\Lambda_{N=2}^{4N_c -4 -2N_f} v^4\\
w_+w_- &=& \zeta\\
\label{CSST}
\left(\prod_{i=1}^{N_f} m_i \right)
y \prod_{i=1}^{N_f}\left(\frac{w_+ - m_i}{w_- - m_i}\right) &=&(-1)^{N_f}
w_+^{2N_c-2}v^2.
\eea
We remark that the solution curve $(\ref{prime-CSSTlike})$ is embedded in
the smooth surface $(\ref{resolved})$ while  the solution curve $(\ref{CSST})$ is embedded in
the singular surface $(\ref{us})$. We repeat that 
the effect of resolving  the singularity
corresponds to pushing  $+4$ charge of O6 through the NS5 branes to
$\pm\infty$, creating new non-dynamical D4 branes  \'a la Hanany-Witten
and thus there are no differnces in the Coulomb branches between
these two models.
\par
Notice that there is a difference in scaling constant between
(\ref{us})  and (5.2) of \cite{csst}. From consideration of the RG
equation:
\bea
\Lambda_{N=2}^{4N_c - 4 - 2N_f} \mu^{2N_c - 2}
= \Lambda_{N=1}^{6N_c - 6 - 2N_f}
\eea
we can write $y$ as
$\Lambda_{N=1}^{3N_c - 3 - N_f}{\mu^{1-N_c}} \tilde{y}$ where $\tilde{y}$
has
dimension 2 as compared with the $y$ in (\ref{CSST}) which has dimension
$2N_c - N_f$. The dimension of $\tilde{y}$ is the same as the one of y
in the equation (5.4) of \cite{csst}
and (\ref{CSST}) becomes almost identical to  (5.4) of \cite{csst} when is
written in terms of $\tilde{y}$.

The extension to $SO(2N_{c} + 1)$ is trivial. In the odd case the
equations (\ref{w_+}) and (\ref{w_-}) become:
\bea
\label{oddw_+}
\prod_{i=1}^{N_f}(w_+ - m_i) y  - P_{1}(w_+) = 0\\
\label{oddw_-}
Q_{1}(w_-) y  - A \prod_{i=1}^{N_f}(w_- -  m_i) = 0
\eea
where
\bea
P_{1}(w_+) =w_+( w_+^{2N_c} + p_1  w_+^{2N_c -1 } + \cdots + p_{2N_c})\\
Q_{1}(w_-) =w_-( w_-^{2N_c} + q_1  w_-^{2N_c -1 } + \cdots + q_{2N_c})
\eea
and  the most general solution for the massive flavor case is of the form
\bea
\label{csstP1}
P_{1}(w_+)& =& w_+^{2N_c + 1 - N_f}\prod_{i=1}^{N_f}(w_+ - \zeta / m_{i})\\
\label{csstQ1}
Q_{1}(w_-)& =& w_-^{2N_c + 1 - N_f}\prod_{i=1}^{N_f} (w_- - \zeta/ m_i)
\eea
which yields
\bea
\label{CSSTlike3}
\left(\prod_{i=1}^{N_f} m_i\right)
y \prod_{i=1}^{N_f}\left(\frac{w_+  - m_i}{w_- - m_i}\right) =(-1)^{N_f}
w_+^{2N_c+1} v^{2}.
\eea

$\bullet$ $Sp(2N_c)$ Case

The situation is similar to $SO$ case. Now  the O6 plane is  described by:
\begin{equation}
x y = \Lambda^{4N_c + 4 - 2N_f} v^{-4}.
\end{equation}
Again the general solution for massive flavors is given by
\bea
xy& =&\Lambda_{N=2}^{4N_c +4 -2N_f}16 \mu^4 (w_+ - w_-)^{-4}\\
w_+w_- &=& \zeta\\
\label{spCSSTlike}
\left(\prod_{i=1}^{N_f} m_i \right)
y \prod_{i=1}^{N_f}\left(\frac{w_+ - m_i}{w_- - m_i}\right) &=&(-1)^{N_f}
w_+^{2N_c}.
\eea
Let us  simplify the spacetime again by resolving singularities at
infinity.
The resolved surface can be given by
\bea
x'y' = \Lambda_{N=2}^{4N_c +4 -2N_f}
\eea
and the new surface maps onto the old surface via the map
$x = x'v^{-2}, y= y' v^{-2}$.
On this new surface, the solution can be described by
\bea
x'y'& =&\Lambda_{N=2}^{4N_c +4 -2N_f}\\
w_+w_- &=& \zeta\\
\label{resspCSSTlike}
\left(\prod_{i=1}^{N_f} m_i \right)
y'\prod_{i=1}^{N_f}\left(\frac{w_+ - m_i}{w_- - m_i}\right) &=&(-1)^{N_f}
w_+^{2N_c+2}.
\eea
If we map this curve to the old surface, then the last equation becomes:
\bea
\label{CSSTlike4}
\left(\prod_{i=1}^{N_f} m_i \right)
y\prod_{i=1}^{N_f}\left(\frac{w_+ - m_i}{w_- - m_i}\right)
&=&(-1)^{N_f}v^{-2}
w_+^{2N_c+2}
\eea
which is the same as (5.27) of \cite{csst} after rescaling of variables.

\section{ Conclusion}
\setcounter{equation}{0}
We have studied brane configurations for
$N=1$ supersymmetric $SO/Sp$ gauge theories and the flavor group was given
by either
infinite D6 branes or semi-infinite D4 branes.
For the case of D6 branes we obtained the results of
\cite{aotaug,aotsept} which were derived from a brane configuration
with an orientifold O4 plane. For the case of semi-infinite D4 branes,
we obtained
 the equations of the Seiberg-Witten curve
as  in \cite{csst} from consideration of
the brane configuration mixed with  simplification of the spacetime via
resolution of the orientifold singularity. While
 these equations
were 
presented with some evidence in \cite{csst}, we derive them rigorously
by fully applying the method of \cite{biksy} and working on
the resolved surface $S'$.
 The simplification of the spacetime via resolution and
other possible solutions deserve further study.

\vspace{2cm}
\centerline{\bf Acknowledgments}

We thank C. Csaki, A. Hanany, A. Karch, D. Kutasov, K. Landsteiner and
Y. Oz for useful discussions and correspondence on relating subjects.
We also thank the referee for thoughtful comments, which resulted in
some corrections and improved readability.


\begin{thebibliography}{99}
\bibitem{gk} A. Giveon and D. Kutasov, hep-th/9802067.
\bibitem{hw} A. Hanany and E. Witten, Nucl. Phys. {\bf B492} (1997) 152,
hep-th/9611230.
\bibitem{bo1}J. de Boer, K. Hori, H. Ooguri, Y. Oz and Z. Yin, Nucl. Phys.
{\bf 493} (1997) 101, hep-th/9612131;
J. de Boer, K. Hori, Y. Oz and Z. Yin, Nucl. Phys. {\bf B502} (1997) 107,
hep-th/9702154.
hep-th/9703051.
\bibitem{egk} S. Elitzur, A. Giveon and D. Kutasov, Phys. Lett. {\bf B400}
(1997) 269, hep-th/9702014.
\bibitem{egkrs}
S. Elitzur, A. Giveon, D. Kutasov, E. Rabinovici and A. Schwimmer, Nucl.
Phys. {\bf B505} (1997) 202,
hep-th/9704104.
\bibitem{eva} N. Evans, C.V. Johnson and A.D. Shapere, Nucl. Phys.
{\bf B505} (1997) 251, hep-th/9703210;
A. Brandhuber, J. Sonnenschein, S. Theisen and S. Yankielowicz,
Nucl. Phys. {\bf B502} (1997) 125, hep-th/9704044;
R. Tatar, Phys. Lett. {\bf B419} (1998) 99, hep-th/9704198.
\bibitem{ov} H. Ooguri and C. Vafa, Nucl. Phys. {\bf B500} (1997) 62,
hep-th/9702180; C. Ahn and K. Oh, Phys. Lett. {\bf B412} (1997) 274,
hep-th/9704061;
C. Ahn and R. Tatar, Phys. Lett. {\bf B413} (1997) 293, hep-th/9705106;
C. Ahn, K. Oh and R. Tatar, hep-th/9707027.
\bibitem{town} P.K. Townsend, Phys. Lett. {\bf B350} (1995) 184,
hep-th/9501068.
\bibitem{w1} E. Witten, Nucl. Phys. {\bf B500} (1997) 3, hep-th/9703166.
\bibitem{lll} K. Landsteiner, E. Lopez and D. Lowe,
Nucl.Phys. {\bf B507} (1997) 197, hep-th/9705199;
A. Brandhuber, J. Sonnenschein, S. Theisen and S. Yankielowicz,
Nucl. Phys. {\bf B504} (1997) 175, hep-th/9705232.
\bibitem{w2}E. Witten, Nucl.Phys. {\bf B507} (1997) 658, hep-th/9706109.
\bibitem{hoo}
K. Hori, H. Ooguri and Y. Oz, Adv.Theor.Math.Phys. {\bf 1} (1998) 1,
hep-th/970682.
\bibitem{biksy} A. Brandhuber, N. Itzhaki, V. Kaplunovsky, J. Sonnenschein
and S. Yankielowicz, Phys.Lett. {\bf B410} (1997) 27, hep-th/9706127.
\bibitem{aotaug} C. Ahn, K. Oh and R. Tatar, hep-th/9708127.
\bibitem{aotsept} C. Ahn, K. Oh and R. Tatar,
Phys. Lett.  {\bf B416} (1998) 75, hep-th/9709096.
107,
\bibitem{ll} K. Landsteiner and E. Lopez, hep-th/9708118.
\bibitem{lll1} K. Landsteiner, E. Lopez and D. Lowe, hep-th/9801002.
\bibitem{bhkl} I. Brunner, A. Hanany, A. Karch and D. Lust, hep-th/9801017.
\bibitem{egkt} S. Elitzur, A. Giveon, D. Kutasov and D. Tsabar,
hep-th/9801020.
\bibitem{csst} C. Csaki, M. Schmaltz, W. Skiba and J. Terning,
hep-th/9801207.
\bibitem{uranga} A.M. Uranga, hep-th/9803054.

\bibitem{witten3} E. Witten, hep-th/9712028.
\bibitem{aps} P.C. Argyres and A.D. Shapere,
Nucl. Phys. {\bf B461} (1996) 437, hep-th/9509175;
E. D'Hoker, I.M. Krichever and D.H. Phong, Nucl. Phys. {\bf B489} (1997)
211, hep-th/9609145.

\bibitem{bhoo} J. de Boer, K. Hori, H. Ooguri and Y. Oz, hep-th/9801060.

\end{thebibliography}
\end{document}